# Extra Dimensions and Atomic Transition Frequencies[*]

Li Zhi-gang (李志刚)[**], Wei-Tou Ni (倪维斗), and Antonio Pulido Patón

Center for Gravitation and Cosmology, Purple Mountain Observatory, Chinese Academy of Sciences, Nanjing 210008

**Abstract**   New unification theories predict Large Extra Dimensions (LEDs). If that is the case, gravity would be stronger at short ranges than what Newtonian gravity predicts. LEDs could also have effects at atomic level. In this paper we propose a new method to constrain the size of "gravity-only" LEDs by analyzing how these LEDs modify the energy of the atomic transitions 1$s$-2$s$ and 2$s$-2$p$ (Lamb shift), for the particular case of the hydrogen and muonium atoms. We estimate these effects by using Bethe's non-relativistic treatment of Lamb shift. For the particular case of three LEDs, which may be a candidate to explain the interaction mechanism of dark matter particles, we have found that current knowledge in atomic spectroscopy could constrain their sizes to be less than 10 μm. Although our contributions do not reach the sensitivity given by SN1987a, they are still slightly better than recent constraints given by Inverse Square Law tests of Eöt-Wash group at Washington University, which gave $R_3$ < 36.6 μm.

**PACC numbers:** 0155, 0450, 2930

## 1. Introduction

The existence of Large Extra Dimensions (LEDs) has received great attention. Recently it has been suggested that the remarkable discrepancy between the Planck scale and electroweak scale, the so-called gauge hierarchy problem in particle physics, could be resolved by introducing appropriate number of spatial extra dimensions with suitable size.

As early as 1990, Antoniadis[1] first suggested TeV-scale LEDs to reexplain supersymmetry breaking. Subsequently others found that LEDs could also be used to lower fundamental physical scales, such as string scale[2], GUT scale[3], seesaw scale[4] and Peccei-Quinn scale.[5]

In 1998 N. Arkani-Hamed, S. Dimopoulos and G. Dvali[6] proposed a model with LEDs (named as ADD model) to explain why gravity is so weak compared to the other three known interactions. Gravity is so weak because it is reduced by its propagation in extra dimensions. In that case gravity could be stronger than expected at short ranges.

It is important to find empirical tests and constraints on LED gravitation theory and theories of gravity in general for the beneficial interaction between theory and experiment. Ni[7] has a recent review on this topic. Adelberger *et al.*[8] has analyzed the empirical constraints on the deviation from inverse square laws. Theories with Ashtekar variables are also of recent interest. Experimental and theoretical tests on these theories together with extra dimension theories are focuses in the literatures.[9-11]

---

[*] Supported by the National Natural Science Foundation of China (Grant No 10475114) and the Foundation of Minor Planets of Purple Mountain Observatory.
[**] Email: zgli@pmo.ac.cn

In the present paper, we discuss how a modified theory of gravity due to "gravity-only" large extra spatial dimensions, could modify the radiation energies of 1s-2s and 2s-2p transitions of hydrogen and muonium atoms. Our estimates are based on Bethe's non-relativistic approximation for calculating the Lamb shift.[12]

We have established upper limits on the sizes $R_3$ and $R_4$ for the cases with three and four LEDs to be $R_3 <$ 10 μm and $R_4 <$ 8.2 nm respectively, by considering recent measurements of the hydrogen and muonium transition frequencies. These constraints are shown in Table. 1. It is worth mentioning that the ADD theory with three LEDs has been considered as an alternative for the interaction mechanism of dark matter described by Bo Qin et al.[13] The ADD model, with three LEDs, can reproduce the inverse-velocity self-interaction cross section of dark matter particles proposed by Firmani et al.[14] to solve the "soft core problem" in the simulation of structure formation of galaxies, clusters and cosmological scales.

In section 2 and 3, we review the ADD model and the empirical constraints on the size of LEDs briefly. In section 4, we calculate the modification to 1s-2s and 2s-2p transition energies of hydrogen and muonium due to extra dimensions. Finally, in section 5 we summarize the main results given in the present paper.

**Table 1.** Upper limits on the compactification radius $R_n$ obtained from our argument and others

|  | transition | $R_3$ | $R_4$ | $R_5$ | $R_6$ |
|---|---|---|---|---|---|
| Hydrogen | 1s-2s | 13 μm | 37 nm | 1.0 nm | 0.1 nm |
|  | 2s-2p | 90 μm | 200 nm | 3.3 nm | 0.3 nm |
| Muonium | 1s-2s | 10 μm | 8.2 nm | 0.1 nm | $6.7 \times 10^{-3}$ nm |
|  | 2s-2p | 30 μm | 17 nm | 0.2 nm | 0.01 nm |
| Inverse Square Law test[24] |  | 36.6 μm | 62 μm |  |  |
| SN1987A[21] |  | 1.14 nm | 0.038 nm | 0.0048 nm | $1.2 \times 10^{-3}$ nm |

## 2. The ADD model

In the ADD model, the Standard Model particles live in a D3-brane with a thickness of about 1 TeV$^{-1}$. The D3-brane is embedded in the $(4 + n)$ dimensional spacetime, where $n$ is the number of spatial extra dimensions, which are compactified on a volume $V_n \sim (2\pi R_n)^n$. In this case, gravitons can propagate freely in the bulk. Using Gaussian law in $(4 + n)$ dimensional spacetime, the gravitational potential between two point particles, of masses $m_1$ and $m_2$, is given by

$$V_{(r)} = \frac{m_1 m_2}{M_{pl(4+n)}^{n+2}} \times \frac{1}{r^{n+1}}, (r << R_n) \quad (1)$$

$$V_{(r)} = \frac{m_1 m_2}{M_{pl(4+n)}^{n+2}} \times \frac{1}{R_n^n \times r}, (r >> R_n) \quad (2)$$

where $M_{pl(4+n)}$ refers to the Planck scale of the $(4 + n)$ dimensional theory and $R_n$ is the size of extra dimensions. Compactification on a $n$-torus has been assumed. The effective 4-dimensional Planck scale can therefore be reinterpreted as $M_{pl}^2 = M_{pl(4+n)}^{2+n} \times R_n^n$. To solve the gauge problem we assume that the $(4 + n)$ dimensional Planck scale and electroweak scale coincide, i.e. $M_{pl(4+n)} \sim m_{ew} \sim$ 1 TeV. To recover the observed 4 dimensional Planck scale we should assume the size of extra dimensions, $R_n$, to be approximately

$$R_n \sim \frac{1}{\pi} 10^{-17+\frac{32}{n}} \left(\frac{1\ \text{Tev}}{m_{ew}}\right)^{1+\frac{2}{n}}\ \text{cm} \qquad (3)$$

The existence of one LED, $R_1 \sim 3 \times 10^{12}$ m, is clearly ruled out by tests of Newtonian gravity over solar system scales. The case of two LEDs, giving $R_2 \sim 0.3$ mm, is also inconsistent with current short-range inverse-square law tests. Nevertheless, theory with more than two LEDs, that give $R_{n>2} \leq 1$ nm are still compatible with the current experimental constraints.

### 3. Empirical constraints on the size of LEDs

Gravity is intimately connected to the geometrical nature of spacetime. It has already been discussed above that LEDs could modify Newtonian gravity at very short distances, below the millimeter range. Inverse-square law tests can therefore give direct evidence for the existence of these entities. High energy colliders can also explore the effects of LEDs in graviton and virtual graviton production and exchange at TeV energy scale. In addition, LEDs can also have some influences in cosmological and astrophysical processes, for example, the cooling of supernovae explosions. In what follows we briefly review these constraints.

**Inverse square law test**  The main obstacle for looking at deviations from Newtonian gravity at short ranges is due to the weakness of those forces. In the last 3 decades, great improvements have been made by various groups in the field of weak force measurements. By testing inverse-square law with a low-frequency torsion oscillator[15, 16], the Eöt-Wash group at Washington University recently put constraints on the size of $n = 3$ and $n = 4$ LEDs compactified on a $n$-torus, to be less than 36.6 μm and 62 μm, respectively, which can be trivially derived by expressing Eq.(18) in Ref.[8] in terms of our Eq.(1) and Eq.(2). We show them in Table 1.

**High energy collider**  In high energy colliders, two kinds of effects due to LEDs are usually considered: a) direct graviton production[17] and b) virtual graviton exchange.[18] Gravitons can be produced directly in processes as $e^+e^-$ and $p^+p^-$, in association with production of photons or jets. The differential cross section of these process depend on the new gravity scale, i.e. $M_{pl(4+n)}$. Virtual gravitons can appear in many Standard Model processes. Virtual gravitons can be traced by measuring deviations of the differential cross section from the Standard Model predictions. These phenomena can be searched for at LEP, TEVATRON, HERA and LHC. Several groups have explored these issues and they put strong constraints on the new Planck scale and thus the size of LEDs in ADD model.[17- 19]

**Cosmology and astrophysics**  Cosmological and astrophysical constraints on LEDs arise from exploration of the production of Kaluza-Klein gravitons coming from Standard Model particles. Over-production of gravitons would increase the density of our universe significantly or even close it. This can give strong constraints on the gravitational scale $M_{pl(4+n)}$ or equivalently $R_n$.[20] Nevertheless this scenario strongly depends on the so-called "normalcy temperature" $T_0$, temperature at which graviton production effectively starts. In that case those constraints could be relaxed by setting larger values of $T_0$.[22] The gravitons produced in supernova would also take energy away and, in this way reduce the observable neutrino emission. By considering the production of Kaluza-Klein gravitons and dilatons via bremsstrahlung processes from the nucleon-nucleon system in SN1987a, Hannestad et al.[21] put stringent constraints on toroidally compactified "gravity-only" LEDs. This scenario depends strongly on the nature of Kaluza-Klein gravitons. These constraints are shown in Table 1.

## 4. Modification to transition energies of hydrogen and muonium due to LEDs

In this section we estimate the energy shift to the 1s-2s and 2s-2p transitions of hydrogen and muonium, when a modified law of gravity at atomic scales is considered. We use Bethe's non-relativistic method for calculating Lamb shift in 1947.[12] A more precise relativistic calculation needs a complete quantum gravity theory which is not known well enough yet. After introducing Bethe's method, we report the detailed calculation of our estimation and put constraints on the size of LEDs.

### 4.1 Bethe's non-relativistic method

In 1947 Bethe applied Kramer's mass renormalization technique to calculate the shift of hydrogen atom energy levels due to the self-energy of electron. A Feynman's diagram describing the self-energy term of electron is shown in Fig.1 in which I stands for intermediate state. Bethe showed that the 2s-2p level shift was approximately 1040 MHz, which remarkably agreed with observations in few percent accuracy.

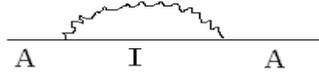

**Fig.1** Feynman's Diagram of the self-energy term of electron in state A

From the classical theory of radiation, where the electron is treated non-relativistic and the electromagnetic field is quantized, the energy shift of electron in state A due to self-energy contributions, is given in Ref.[23] by

$$\Delta E_A = \frac{e^2}{6\pi^2 \varepsilon_0 \hbar m_e^2 c^3} \sum_I \int dE_\gamma \frac{E_\gamma |(\mathbf{p})_{IA}|^2}{E_A - E_I - E_\gamma} \qquad (4)$$

where $m_e$ and $\mathbf{p}$ is the mass and momentum of electron, respectively, and $E_\gamma$ is the energy of emitted and/or absorbed photon shown in Fig.1. The sum is performed all over the intermediate states I. The result is obviously linearly divergent. The shift of the kinetic energy in the case of a free electron with momentum $\mathbf{p}$ is given by

$$\Delta E_{free} = -\frac{e^2 \mathbf{p}^2}{6\pi^2 \varepsilon_0 \hbar m_e^2 c^3} \int_0^{E_\gamma^{max}} dE_\gamma \equiv C \cdot \mathbf{p}^2 \qquad (5)$$

where $E_\gamma^{max}$ is the cut-off energy of the interacting photon. In a non-relativistic approximation this energy cut-off is set to be the electron rest mass, i.e. $m_e c^2$.

The energy term defined by Eq.(5) cannot be "observed" separately from the electron kinetic energy in state A. Therefore the main idea of mass renormalization will then consist of subtracting Eq.(5) from Eq.(4). The observable energy shift is then given by

$$\Delta E_A^{(obs)} = -\frac{e^2}{6\pi^2 \varepsilon_0 \hbar m_e^2 c^3} \ln \frac{E_\gamma^{max}}{<E_I - E_A>_{Av}} \sum_I |(\mathbf{p})_{IA}|^2 (E_I - E_A) \qquad (6)$$

The average excitation energy $<E_I - E_A>_{Av}$ over all intermediate states is defined and calculated numerically in Ref.[12, 24]. The sum term can also be easily calculated, giving

$$\sum_I |(\mathbf{p})_{IA}|^2 (E_I - E_A) = -\frac{\text{A}^2}{2} \int d^3\mathbf{x} |\Psi_A|^2 \nabla^2 V(\mathbf{x}) \qquad (7)$$

where $V(x)$ is the potential part of the unperturbed Hamiltonian and $\Psi_A$ is wave function of electron in state A. Taking all of the discussions above into account, the observable energy shift can be written as

$$\Delta E_A^{(obs)} = -\frac{e^2 \text{A}}{12\pi^2 \varepsilon_0 m_e^2 c^3} \ln \frac{E_\gamma^{max}}{<E_I - E_A>_{Av}} \int d^3\mathbf{x} |\Psi_A|^2 \nabla^2 V(\mathbf{x}) \qquad (8)$$

Taking the electron's wave function of hydrogen as $\Psi_A$ and $V(x)$ the Coulomb potential, one obtains $\Delta v_{2s-2p}^{obs}$ = 1040 MHz. This result is in excellent agreement (2%) with the experimental value of 1057.85 MHz.

As a conclusion we can estimate the leading order of the energy shift caused by modified gravity using this non-relativistic approximation, as we will do in what follows.

**4.2 Energy shift due to a modified Newtonian law of gravity**

In this section we estimate the energy shift to the 1s-2s and 2s-2p transitions for the particular case of hydrogen and muonium atoms due to a modified gravitational law at short distances. As mentioned above, the gravitational interaction could be stronger as expected at atomic or subatomic scale due to the presence of LEDs. Because this effect is still much weaker than those due to the electromagnetic interaction, the wave function of the electron is unaffected to first order approximation. We can also treat gravity as a small perturbation to the Coulomb potential. Inserting the gravitational potential Eq.(1) and Eq.(2) into Eq.(8), we can write the shift of energy levels by modified gravity as

$$\Delta E_A^{(obs)} = -\frac{e^2 \text{A}}{12\pi^2 \varepsilon_0 m_e^2 c^3} \ln \frac{E_\gamma^{max}}{<E_I - E_A>_{Av}} \int d^3\mathbf{x} |\Psi_A|^2 \nabla^2 V_g(\mathbf{x}) \qquad (9)$$

Because of the finite size of the proton we cut off the integration volume at the proton radius, for the hydrogen atom case, that is $r_c = r_p$ = 0.875 fm where $r_c$ is the cut-off distance. In the case of muonium, the electron's wave function determined by the electromagnetic interaction, should be valid until, approximately, $10^{-17}$ m. At that distance the corrections due to weak interaction become important. We should also remember that the gravitational potential due to LEDs, in the ADD model, should switch off theoretically at distances below the electroweak distance scale ~ $10^{-18}$ m (1 TeV$^{-1}$). Taking these into account, we can safely take $r_c = 10^{-17}$ m as the cut-off distance when performing the integration in Eq.(9).

The average excitation energy in the logarithmic factor in Eq.(9), for the 2s state, was numerically calculated by Bethe *et al.*[24] They found out that the major contribution (about 97 percent) arised from the continuous spectrum, while the discrete part only contributed less than 3 percent. Bethe and colleagues reported a value of 16.646 ± 0.007 Ry, where Ry (13.6 eV) is the ionization energy of the ground state of hydrogen. This value is much smaller than the electron's rest mass (0.511 MeV) and does not contribute appreciably to the logarithmic factor. In this way, we have taken the average excitation energy for the 1s state to be approximately the value numerically calculated in Ref.[24] for the 2s state, that is $<E_I - E_{1s}>_{Av} \sim <E_I - E_{2s}>_{Av}$. This will result in an uncertainty of less than 20 percent in our constraints.

Then by performing the integrations in Eq.(9) we obtain the energy shifts by modified gravity due to LEDs. We show them in Table 2, in which $B_n$ is defined as $(R_n / \rho_B)^n$ and $\rho_B = 5.29$ nm is the Bohr radius of hydrogen. H and M stand for hydrogen and muonium, respectively.

Table 2.  The energy shift of hydrogen and muonium due to modified gravity.

|   | State A | $n = 3$ | $n = 4$ | $n = 5$ | $n = 6$ |
|---|---|---|---|---|---|
| H | 1s | $1.3 \times 10^{-29} B_3$ eV | $1.0 \times 10^{-24} B_4$ eV | $7.4 \times 10^{-20} B_5$ eV | $5.1 \times 10^{-15} B_6$ eV |
| H | 2s | $1.7 \times 10^{-30} B_3$ eV | $1.3 \times 10^{-25} B_4$ eV | $9.3 \times 10^{-21} B_5$ eV | $6.4 \times 10^{-16} B_6$ eV |
| M | 1s | $6.8 \times 10^{-24} B_3$ eV | $8 \times 10^{-17} B_4$ eV | $1 \times 10^{-9} B_5$ eV | $1.1 \times 10^{-2} B_6$ eV |
| M | 2s | $0.85 \times 10^{-24} B_3$ eV | $1 \times 10^{-17} B_4$ eV | $1.3 \times 10^{-10} B_5$ eV | $1.4 \times 10^{-3} B_6$ eV |

**4.3 Constraints on the size of LEDs from hydrogen atom spectroscopy**

Measurements of hydrogen transition frequencies have been improved during the last two decades. By using a laser-cooled cesium atom clock and with several other improvements, Niering et al.[25] have achieved the most accurate measurement of hydrogen 1s-2s transition frequency. They report a value of 2466 061 413 187.103 (46) KHz ($1.9 \times 10^{-14}$), which is commonly used in the current adjustment for fundamental physical constants.[26] For the 2s-2p level splitting, the Lamb shift of hydrogen, a result of 1057851.4 (1.9) KHz has been reported by Pal'chikov et al.[27] in 1985, which has the best precision up to date. The theoretical calculation is limited by the mean square charge radius of proton.

With these precise experimental data in hand, we can put stringent constraints on LEDs. We show these constraints in Table 1, together with the constraints from inverse-square law tests and astrophysical phenomena (SN1987a).

**4.4 Constraints on the size of LEDs from muonium atom spectroscopy**

The muonium atom, consisting of two simplest leptons, a positive muon and an electron, was first discovered by Vernon W. Hughes in 1960.[28] Muonium spectroscopy has received a great impulse since then. The Doppler-free excitation of 1s-2s muonium transition was achieved at KEK[29] and independently at RAL.[30] The latest experiments at RAL give $\Delta v_{1s\text{-}2s}$ (exp) = 2 455 528 941.0(9.8) MHz.[31] While the accuracy of the theoretical value, $\Delta v_{1s\text{-}2s}$ (theo) = 2 455 528 935.4(1.4) MHz, is limited mainly by the uncertainty of the muon's rest mass.

Contrary to the highly precise value for 1s-2s transition frequency, the measured value for 2s-2p (Lamb shift) transition frequency is not so accurate. The experiments at LAMPF[32] and at TRIUMPF[33] give 1042(+21)(-23) MHz and 1070(+12)(-15), respectively. Constraints on the size of LEDs estimated from these measurements are also shown in Table 1.

There are several plans for muon research in the future, such as the J-PARC facility in Japan and EURISOL facility in Europe.[34] These plans aim at more accurate values for transition frequencies of muonium, and in this way our constraints will be improved.

**5. Conclusion**

LEDs have received great attention because of its value in solving the gauge hierarchy problem and lowering other fundamental physical scales. We have estimated the effects of a modified gravitational interaction due to LEDs in the atomic transition of hydrogen and muonium atoms. We have established stringent constraints on the compactification radius of the LEDs based on ADD theoretical model. In particular, for the special cases of three and four spatial LEDs compactified on a torus, our constraints, $R_3 < 10$ μm and $R_4 < 8.2$ nm, are slightly better than the recent constraints, $R_3 < 36$ μm and $R_4 < 62$ μm, established with a torsion balance experiment to test the inverse-square law of gravity at Washington.[8,16] Of course our constraints and the constraints from inverse-square law tests are both not as stringent as those from astrophysics and cosmology which depend on the nature of quantum theory of gravity. The subsequent relativistic consideration and the numerical calculation of average excitation energy are needed.

**References**

[1] Antoniadis I 1990 *Phys. Lett.* B 246 377
[2] Witten E 1996 *Nucl. Phys.* B 471 135
[3] Lykken J D 1996 *Phys. Rev.* D 54 3693
[4] Dienes K R, Dudas E and Gherghetta T 1999 *Nucl. Phys.* B 557 25
[5] Chang S, Tazawa S and Yamaguchi M 2000 *Phys. Rev.* D 61 084005; Dienes K R, Dudas E and Gherghetta T 2000 *Phys. Rev.* D 62 105023
[6] Arkani-Hamed N, Dimopoulos S, and Dvali G 1998 *Phys. Lett.* B 429 263
[7] Ni W-T 2005 *Int. J. Mod. Phys.* D 14 901
[8] Adelberger E G, Heckel B R, Hoedl S, Hoyle C D and Kaoner D J 2007 *Phys. Rev. Lett.* 98 131104
[9] Tang Z M 2000 *Chin. Phys.* 9 0881
[10] Li J L, Wu Y B and Li L 2002 *Chin. Phys.* 11 0327
[11] Tao B X, Ji Sh Y and Li F Q 2004 *Chin. Phys.* 13 1830
[12] Bethe H A 1947 *Phys. Rev.* 72 4
[13] Qin B, Pen U L and Silk J 2005 *News in Nature* [arXiv: astro-ph/0508572]
[14] Firmani C, D'Onghia E, Avila-Reese V, Chincarini G, and Hernandez X 2000 *Mon. Not. R. Astron. Soc.* 315 L29
[15] Hoyle C D, Kapner D J, Hekel B R, Adelberger E G, Gundlach J H, Schmidt U, and Swanson H E 2004 *Phys. Rev.* D 70 042004
[16] Kapner D J, Cook T S, Adelberger E G, Gundlack J H, Heckel B R, Hoyle C D and Swanson H E 2007 *Phys. Rev. Lett.* 98 021101
[17] Vacavant L and Hinchlitte I 2001 *J. Phys.* G 27 1839; Giudice G F, Rattazzi R and Wells J D 1999 *Nucl. Phys.* B 544 3; Han T, Lykken J D and Zhang R J 1999 *Phys. Rev.* D 59 105006; Mirabelli E A, Perelstein M and Peskin M E 1999 *Phys. Rev. Lett.* 82 2236; Atwood D, Bar-Shalom S and Soni A 2000 *Phys. Rev.* D 61 116011; Cullen S, Perelstein M and Peskin M E 2000 *Phys. Rev.* D 62 055012
[18] Giudice C F, Rattazzi R and Well J D 1999 *Nucl. Phys.* B 544 3; Atwood D, Bar-Shalom S and Soni A 2000 *Phys. Rev.* D 61 054003; Dvergsnes E, Osland P and Ozturk N 2003 *Phys. Rev.* D 67 074003
[19] Landsberg G 2004 to appear in *Proceedings of SLAC Summer Institute*
[20] Hall L J and Smith D 1999 *Phys. Rev.* D 60 085008
[21] Hannestad S and Raffelt G G 2003 *Phys. Rev.* D 67 125008
[22] Macesanu C and Trodden M 2004 *Phys. Rev.* D 71 051902
[23] Sakurai J J 1978 *Advanced Quantum Mechanics* (Philippines: Addison-Wesley Publishing Company) p70
[24] Bethe H A, Brown L M and Stehn J R 1950 *Phys. Rev.* 77 370

[25] Niering M, Holzwarth R, Reichert J, Pokasov P, Udem Th, Weitz M and Hänsch T W 2000 *Phys. Rev. Lett.* 84 5496
[26] Mohr P J and Taylor B N 2005 *Rev. Mod. Phys.* 77 1
[27] Pal'chikov V G, Sokolov Y L and Yakovlev V P 1985 *Metrologia* **21** 99
[28] Hughes V W, McColm D W and Ziock K 1960 *Phys. Rev. Lett.* 5 63
[29] Steven Chu, Mills A P Jr, Yodh A G, Nagamine K, Miyake Y and Kuga T 1988 *Phys. Rev. Lett.* 60 101
[30] Maas F E, Lander G H, Longfield M J, Langridge S, Mannix D, E Lidstrom et al. 1994 *Phys. Lett.* A 187 247
[31] Meyer V, Bagaev S N, Baird P E G, Bakule P, Boshier M G, Breitruck A et al. 2000 *Phys. Rev. Lett.* 84 1136
[32] Woodle K A, Badertscher A, Hughes V W, Lu D C, Ritter M W, Gladisch M, Orth H, zu Putlitz G, Eckhause M, Kane J and Mariam F G 1990 *Phys. Rev.* A 41 93
[33] Oram C J, Bailey J M, Schmor P W, Fry C A, Kiefl R F, Warren J B, Marshall G M and Olin A 1984 *Phys. Rev. Lett.* 52 910
[34] Jungmann K P 2003 *Proceedings of the Memorial Symposium in honor of Vernon Willard Hughes*